\documentstyle[psfig,epsfig]{mn}

\newif\ifAMStwofonts

\def\simlt{\lower.5ex\hbox{$\; \buildrel < \over \sim \;$}}
\def\simgt{\lower.5ex\hbox{$\; \buildrel > \over \sim \;$}}
\def\etal{{\it et al.}\ }

\ifoldfss
  \ifCUPmtlplainloaded \else
    \NewTextAlphabet{textbfit} {cmbxti10} {}
    \NewTextAlphabet{textbfss} {cmssbx10} {}
    \NewMathAlphabet{mathbfit} {cmbxti10} {} 
    \NewMathAlphabet{mathbfss} {cmssbx10} {} 
  \fi
  \ifAMStwofonts
    \ifCUPmtlplainloaded \else
      \NewSymbolFont{upmath} {eurm10}
      \NewSymbolFont{AMSa} {msam10}
      \NewMathSymbol{\upi}     {0}{upmath}{19}
      \NewMathSymbol{\umu}     {0}{upmath}{16}
      \NewMathSymbol{\upartial}{0}{upmath}{40}
      \NewMathSymbol{\leqslant}{3}{AMSa}{36}
      \NewMathSymbol{\geqslant}{3}{AMSa}{3E}

       \let\le=\leqslant
       \let\ge=\geqslant
    \fi
  \fi
\fi 

\ifnfssone
  \newmathalphabet{\mathit}
  \addtoversion{normal}{\mathit}{cmr}{m}{it}
  \addtoversion{bold}{\mathit}{cmr}{bx}{it}
  \newmathalphabet{\mathbfit} 
  \addtoversion{normal}{\mathbfit}{cmr}{bx}{it}
  \addtoversion{bold}{\mathbfit}{cmr}{bx}{it}
  \newmathalphabet{\mathbfss} 
  \addtoversion{normal}{\mathbfss}{cmss}{bx}{n}
  \addtoversion{bold}{\mathbfss}{cmss}{bx}{n}
  \ifAMStwofonts
    \ifCUPmtlplainloaded \else
      \UseAMStwoboldmath
      \makeatletter
      \new@mathgroup\upmath@group
      \define@mathgroup\mv@normal\upmath@group{eur}{m}{n}
      \define@mathgroup\mv@bold\upmath@group{eur}{b}{n}
      \edef\UPM{\hexnumber\upmath@group}
      \new@mathgroup\amsa@group
      \define@mathgroup\mv@normal\amsa@group{msa}{m}{n}
      \define@mathgroup\mv@bold\amsa@group{msa}{m}{n}
      \edef\AMSa{\hexnumber\amsa@group}
      \makeatother
      \mathchardef\upi="0\UPM19
      \mathchardef\umu="0\UPM16
      \mathchardef\upartial="0\UPM40
      \mathchardef\leqslant="3\AMSa36
      \mathchardef\geqslant="3\AMSa3E

       \let\le=\leqslant
       \let\ge=\geqslant
    \fi
  \fi
\fi 

\ifnfsstwo
  \DeclareMathAlphabet{\mathbfit}{OT1}{cmr}{bx}{it}
  \SetMathAlphabet\mathbfit{bold}{OT1}{cmr}{bx}{it}
  \DeclareMathAlphabet{\mathbfss}{OT1}{cmss}{bx}{n}
  \SetMathAlphabet\mathbfss{bold}{OT1}{cmss}{bx}{n}
  \ifAMStwofonts
    \ifCUPmtlplainloaded \else
      \DeclareSymbolFont{UPM}{U}{eur}{m}{n}
      \SetSymbolFont{UPM}{bold}{U}{eur}{b}{n}
      \DeclareSymbolFont{AMSa}{U}{msa}{m}{n}
      \DeclareMathSymbol{\upi}{0}{UPM}{"19}
      \DeclareMathSymbol{\umu}{0}{UPM}{"16}
      \DeclareMathSymbol{\upartial}{0}{UPM}{"40}
      \DeclareMathSymbol{\leqslant}{3}{AMSa}{"36}
      \DeclareMathSymbol{\geqslant}{3}{AMSa}{"3E}

       \let\le=\leqslant
       \let\ge=\geqslant
    \fi
  \fi
\fi 

\ifCUPmtlplainloaded \else
  \ifAMStwofonts \else 
    \def\upi{\pi}
    \def\umu{\mu}
    \def\upartial{\partial}
  \fi
\fi

\title{Image Simulation with Shapelets}
\author[R.~Massey \etal]
{Richard~Massey,$^1$\thanks{E-mail: {\tt rjm@ast.cam.ac.uk}}
Alexandre~Refregier,$^{2,1,3}$ Christopher~J.~Conselice$^3$ and
\newauthor David~J.~Bacon$^4$\\
$^1$ Institute of Astronomy, Madingley Road, Cambridge CB3 OHA. \\
$^2$ Service d'Astrophysique, B\^{a}t. 709, CEA Saclay, F-91191 Gif sur
     Yvette, France.\\
$^3$ California Institute of Technology, 1201 E. California Blvd., Pasadena,
     CA 91125, USA.\\
$^4$ Institute for Astronomy, Blackford Hill, Edinburgh EH9 3HJ.}

\date{Accepted 28 October 2003; in original form 21 January 2003.}

\pagerange{\pageref{firstpage}--\pageref{lastpage}}
\pubyear{2003}

\begin{document}

\maketitle

\label{firstpage}

\begin{abstract} We present a method to simulate deep sky images, including
realistic galaxy morphologies and telescope characteristics. To achieve a wide
diversity of simulated galaxy morphologies, we first use the shapelets
formalism to parametrize the shapes of all objects in the Hubble Deep Fields.
We measure this distribution of real galaxy morphologies in shapelet parameter
space, then resample it to generate a new population of objects. These
simulated galaxies can contain spiral arms, bars, discs, arbitrary radial
profiles and even dust lanes or knots. To create a final image, we also model
observational effects, including noise, pixellisation, astrometric distortions
and a Point-Spread Function. We demonstrate that they are realistic by showing
that simulated and real data have consistent distributions of morphology
diagnostics: including galaxy size, ellipticity, concentration and asymmetry
statistics. Sample images are made available on the world wide web. These
simulations are useful to develop and calibrate precision image analysis
techniques for photometry, astrometry, and shape measurement. They can also be
used to assess the sensitivity  of future telescopes and surveys for
applications such as supernova searches, microlensing, proper motions, and weak
gravitational lensing. \end{abstract}


\section{Introduction} \label{intro}

As astronomical surveys are growing in size and scope, so image analysis
methods are increasing in complexity and accuracy. In order to calibrate these
new methods, it is essential to have a large sample of images containing
objects whose properties are already known. Since real data is subject to the
uncertainties of observational noise, telescope aberration and seeing, several
packages have been developed to manufacture artificial images ({\it e.g.}~{\tt
Skymaker} (see Erben \etal2001) or {\tt artdata} in {\tt IRAF} (Tody 1993)).
The accuracy of image analysis methods can then be assessed by comparing their
output to the known input image properties that were specified before the
addition of such observational effects.

The image simulation packages currently available are particularly
valuable for imitating deep ground-based data. However, they limit
themselves to a representation of galaxies as parametric forms
such as symmetric de Vaucouleurs or exponential profiles. Deep
space-based images, on the other hand, contain many irregular or
asymmetrical galaxies with complex resolved features such as
spiral arms, mergers and dust lanes. One possibility for
simulating space images, utilised by Bouwens, Broadhurst \& Silk
(1998), is to repeatedly reuse well-resolved galaxies from the
Hubble Deep Fields (HDFs; Williams \etal1996, 1998). However, this
restricts us to morphology templates from a relatively bright and
nearby sample. Fainter galaxies cannot be used because they have
been significantly contaminated with background noise.
Consequently, the morphological properties of the faint galaxy
population are not fairly represented. This method also faces the
difficulty that the same real galaxies must be reused many times
within one simulation. Although the HDFs are indeed very deep
($I_{F814W}$=27.60 at 10$\sigma$, Williams \etal1996), they only
cover a small area ($\sim$6 square arcminutes each) and contain a
finite number of galaxies. Even if we were to source our real
galaxies from larger surveys such as the Groth strip (Groth
\etal1994) or the Medium Deep Survey (Ratnatunga, Griffiths \&
Ostrander 1999), we would still face the difficulty of using
particular real galaxies many times in a large simulation.

In this paper, we present a method for simulating deep images that contain
genuinely unique objects, yet replicate the morphological distribution of
galaxies in the HDF at all depths. This method has the advantage of allowing us
to simulate arbitrarily large, deep surveys with no repetition of galaxy
shapes. It also allows us to know accurately the intrinsic properties of each
galaxy, before adding telescope-specific noise properties, systematic effects
and convolution with a Point-Spread Function (PSF).

Our method is to decompose all objects in the HDFs into shapelet
parametrizations, following the formalism introduced by Refregier (2003,
hereafter Shapelets~I) and Refregier \& Bacon (2003; hereafter Shapelets~II).
Using just a few coefficients, these can completely quantify the shape
properties of all galaxies, including spiral arms, bars and arbitrary radial
profiles. We then model their distribution of shapelet coefficients, and draw
from this probability distribution new sets of shapelet coefficients,
representing new galaxies. In particular, we take into account the covariance
between shapelet coefficients so that, for example, shapes depend upon
magnitude and size ({\it e.g.}~faint galaxies appear more irregular than bright
ones). In this method, we therefore do not input any model of physical
morphology or evolution. Rather, we exclusively use the measured statistics of
the shapelet coefficient distributions from a real galaxy sample, as a function
of magnitude and size. The new galaxy images can then be analytically convolved
with any PSF, pixellated, and given an appropriate amount of noise for any
exposure time down to the depth of the HDF.

These simulations have several significant applications. We can
use them to calibrate the effectiveness of image analysis and
detection methods such as {\tt SExtractor} (Bertin \& Arnouts
1996), {\tt imcat} (Kaiser, Squire \& Broadhurst 1995), {\tt
GIM2D} (Simard 1998), {\tt GALFIT} (Peng \etal2002) and wavelet
routines ({\it e.g.}~Meyer 1993). By examining the errors on shape
measurement at various signal-to-noise levels of galaxy detection,
we can also predict the accuracy of future experiments requiring
accurate shape measurement. An example of this for space-based
cosmic shear surveys is presented in Massey \etal(2003).

This paper is organized as follows. In \S\ref{mk_decomp} we give a brief
overview of the shapelet formalism and describe how the HDF galaxies are
modelled using shapelet coefficients. In \S\ref{adddisadd}, we show how the
properties of the shapelet basis functions make them eminently suitable for
this method. In \S\ref{method} we discuss the means by which we recover a
smooth probability distribution of galaxy morphologies in shapelet parameter
space. In \S\ref{results} we generate new galaxies by resampling the
distribution. We then add observational noise and show an example of the final
simulated images.

We then demonstrate that the simulations do indeed have similar properties to
the HDFs. For this purpose, we consider in \S\ref{tests} commonly used
quantifiers for galaxy morphology. We find good agreement between simulations
and the real HDF galaxies for measures such as the size-magnitude distribution,
ellipticity, concentration, asymmetry and clumpiness indices ({\it
e.g.}~Bershady \etal 2000, Conselice \etal2000a). It is this agreement which is
the final justification of our shapelet-based simulation method. We compare our
method to others in \S\ref{compare} and summarise our findings in
\S\ref{conclusions}. Sample images may be downloaded from {\tt
http://www.ast.cam.ac.uk/$\sim$rjm/shapelets}.

\section{Shapelet source catalogue} \label{mk_decomp}

In this section, we describe the detection of HDF galaxies and their modelling
as shapelets. This procedure creates a parametrized catalogue of real galaxy
morphologies, which we will require later.

\subsection{Source detection} \label{catgen}

\begin{figure}
 \psfig{figure=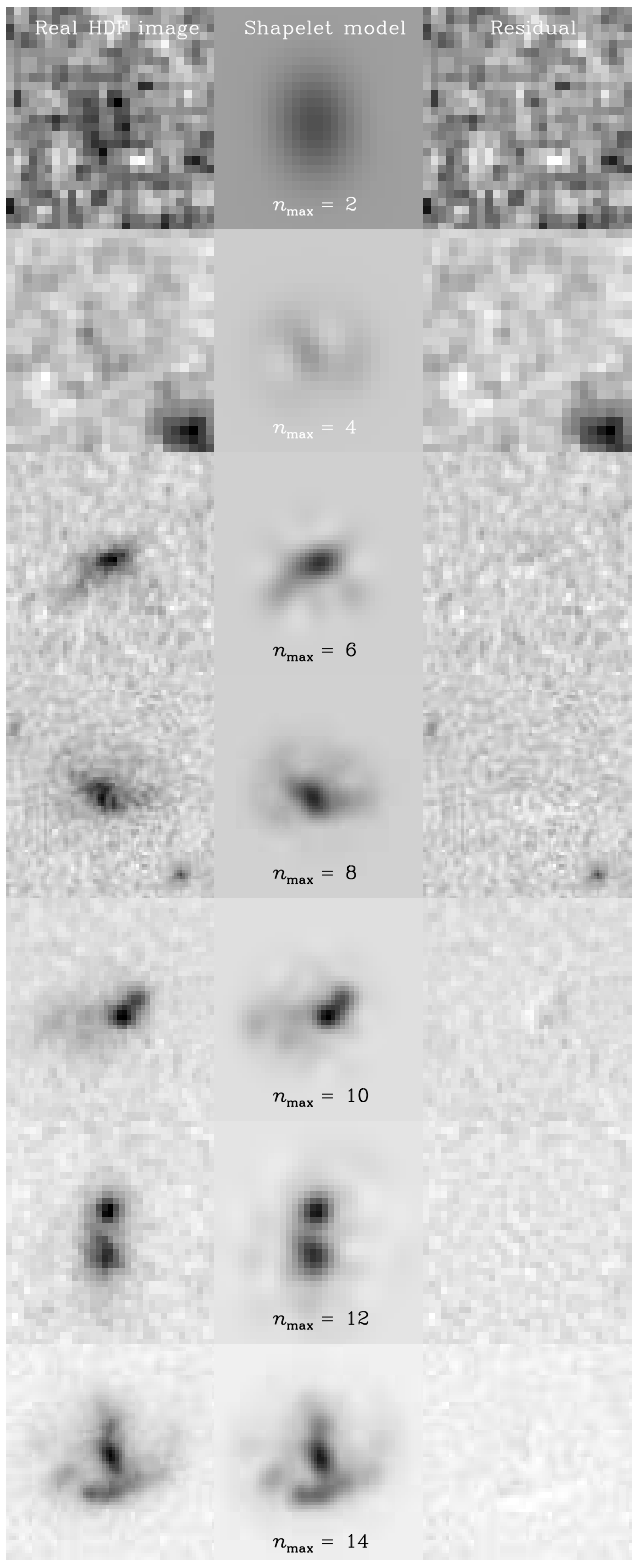,width=83mm}
 \caption{Shapelet modelling of a selection of HDF $I$-band galaxies. Higher
 S/N galaxies typically require more shapelet coefficients so we display a
 variety of source galaxies, noting the shapelet $n_{\rm max}$ required to reach
 a reconstruction with $\chi^2_r=1$. In all cases, the first column
 shows the original HDF image; the middle column shows the shapelet model; the
 right column shows the residual. The image size and colour scale is different
 for each row. \label{fig:gals}}
\end{figure}

Objects are initially detected using {\tt SExtractor} (Bertin \& Arnouts 1996)
upon the HDF $I$-band (F814W) images, together with the pixel weight maps
outputted by {\tt DRIZZLE} (Fruchter \& Hook 2002). The convolution mask and
detection parameters were adapted from those used by Williams \etal(1996). In
particular, we use a comparatively low S/N detection threshold, ${\mathrm
DETECT\_THRESH}$, of 1.3. This affords recovery of faint galaxies and minimizes
incompleteness, at the expense of many false-positive `detections' of noise,
which need to be flagged and filtered out later (see \S\ref{decomp}). Stars
with ${\mathrm CLASS\_STAR}>97$\% are immediately discarded, as we wish to
model only galaxies. The image is then segmented into small square `postage
stamp' regions around the remaining galaxies. The sizes of these regions are
set to $(3\times{\mathrm A\_IMAGE}+5)$ pixels square, where ${\mathrm
A\_IMAGE}$ is a measure of the galaxy's major axis provided by {\tt
SExtractor}. This area is slightly smaller than those illustrated in
figure~\ref{fig:gals}; it is compact enough to be computationally efficient,
but large enough to ensure that the shapelet basis functions are close to zero
at the boundaries of the image.

This prescription conveniently leaves a border of sky background and noise
around the edge of each image. We use all of the border pixels that
do not belong to any other object in the {\tt SExtractor} catalogue to
locally renormalise the pixel weight map. As noticed by Williams
\etal(1996), the inverse variance map outputted during the data
reduction of the HDF systematically overestimates the noise by a
factor of a few. This bias also varies as a function of position
around the image. While {\tt SExtractor} requires only relative
weights between pixels, and is thus unaffected by this bias, we
need to calibrate the absolute value of noise for the shapelet
decomposition.

\subsection{Shapelet modelling} \label{decomp}

Shapelets are a complete, orthonormal set of 2D basis functions. A
linear combination of these functions can be used to model any
image, in a similar way to Fourier or wavelet synthesis. The
shapelet decomposition is particularly efficient for images
localised in space, such as those of individual galaxies. The
formalism was first introduced in Shapelets~I, and a related
method has also been independently suggested by Bernstein \&
Jarvis (2002).

For the polar shapelet analysis, the surface brightness $f({\mathbf x})$ of an
object can be written as
\begin{equation}
\label{eq:decompose_2d}
  f({\mathbf x}) = \sum_{n=0}^{\infty} \sum_{m=-n}^{n}
  a_{nm} \chi_{nm}({\mathbf x}-{\mathbf x}_c;\beta)~,
\end{equation}
\noindent where $\beta$ is a scale parameter, and ${\mathbf x}_c$ is the
position of the centre of the basis functions. Only combinations of $n$ and $m$
where both are even or both are odd should be included in this summation. The
basis functions $\chi_{nm}$, expressed in their polar separable form, are
given by

\begin{eqnarray}
\chi_{n,m}(r,\theta;\beta) = ~~~~~~~~~~~~~~~~~~~~~~~~~~~~~~~~~~~~~~~~~~~~~~~~~~~ \nonumber \\
  \frac{(-1)^{\frac{n-|m|}{2}}}{\beta^{|m|+1}\sqrt{\pi}} 
  \left[\frac{\left(\frac{n-|m|}{2}\right)!}{\left(\frac{n+|m|}{2}\right)!}\right]^{\frac{1}{2}}
  r^{|m|}
  L_\frac{n-|m|}{2}^{|m|} \left(\frac{r^2}{\beta^2}\right)
  e^\frac{-r^{2}}{2\beta^2}
  e^{im\theta} ,
\end{eqnarray}
\noindent where $L(x)$ are the Laguerre polynomials (see
{\it e.g.}~Boas 1983). The index $n$ describes the 
radial oscillations and the index $m$ describes the order of
rotational symmetry. Orthonormality ensures that the shapelet
coefficients $a_{nm}$ are given by
\begin{equation} \label{eqn:lindecompp}
a_{n,m} = \int_0^{2\pi} d\theta \int_0^\infty r~dr~f(r,\theta)
~\chi_{n,m}(r,\theta;\beta)~.
\end{equation}
\noindent These are Gaussian-weighted multipole moments of the
surface brightness, familiar in several branches of astronomy.

For reasonable choices of the centroid ${\mathbf x}_c$ and scale size $\beta$,
the galaxy shape information is contained within only the first few shapelet
coefficients. The series in equation~(\ref{eq:decompose_2d}) can then be
truncated at some finite order $n_{\rm max}$. In order to make good choices for
${\mathbf x}_c$, $\beta$ and $n_{\rm max}$, we first define $\chi^2_r$ as the
difference between the original and reconstructed image, renormalised with
respect to the local noise level. We then attempt to find the values of 
${\mathbf x}_c$ and $\beta$ which achieve $\chi^2_r=1$ with the fewest possible
shapelet coefficients, or minimum $n_{\rm max}$. Shapelet coefficients with
higher $n$ can be discarded and the shapelet model will still be consistent
with the data.

A practical algorithm to perform this optimisation by iteratively exploring
$\{{\mathbf x}_c$, $\beta$, $n_{\rm max}\}$ space is described in Massey \&
Refregier (2003). The algorithm creates a catalogue of optimised shapelet
decompositions for $\sim500$ objects per square arcminute in the HDFs. However,
this represents only 81\% of the `objects' detected by {\tt SExtractor}.
Approximately two dozen of the brightest of these galaxies require a
decomposition with $n_{\rm max}>15$ to achieve $\chi^2_r\le1$. To reduce the
dimensionality in later analysis, these parametrizations are truncated at this
point regardless, with ${\mathbf x}_c$ and $\beta$ chosen to give the best
possible, if slightly imperfect, shapelet fit. The algorithm also fails to
converge to fits with $\chi^2_r\le1$ for a further 42 galaxies in close pairs,
as identified by the {\tt SExtractor} segmentation map; 36 galaxies because of
their proximity to bright stars or the edge of the image; and 60 more objects
across the HDFs (about 10\% of all {\tt SExtractor} detections), which are
mainly false detections of noise due to the low S/N detection threshold set in
section \S\ref{catgen}. Note that the number of decompositions which fail due
to contamination from a near neighbour is roughly independent of magnitude.
Indeed, the slope of the number counts for galaxy pair members is within
$1\sigma$ of that for all the galaxies in the HDF: therefore this particular
effect should not introduce any bias.

Figure~\ref{fig:gals} displays a selection of HDF galaxies at various S/N
levels, and their shapelet reconstructions (see also Shapelets~I, figures~3 and
4). Faint galaxies typically require an $n_{\rm max}$ of only 2, 3 or 4, while
brighter, larger objects require an increasing number of shapelet coefficients
to model their greater degree of detail. The right-hand column of
figure~\ref{fig:gals} shows the reconstruction residuals, which are consistent
with noise even for irregular galaxy morphologies.

\subsection{Treatment of the PSF}\label{deconv}

During the modelling of galaxy shapes, we must in general account for the PSF
of the WFPC2 camera that has smeared the HDF images. Since our objective here
is to simulate only HST images, we do not apply any correction. The PSF will be
naturally contained within the shapelet parametrization of the galaxy images
and these are intentionally left unaltered. When we create simulated images,
they will automatically have been smeared by the WFPC2 PSF: effectively
circularised on average, because of the random reorientation of the new
galaxies.

However,~for~other~applications~it~may~be~desir\-
able~to~simulate~observations~from~other~telescopes
such~as~the~JWST~({\tt http://www.stsci.edu/ngst/}), SNAP ({\tt
http://snap.lbl.gov/}) or GAIA ({\tt http://astro.
esa.int/gaia/}). It would then be necessary to take account of
their different instrumental properties. The ideal way to do this
would be to deconvolve HDF galaxies from the WFPC2 PSF
analytically in shapelet space (see Shapelets~II \S3), and then
to reconvolve simulated galaxies with a new PSF at the end.
Unfortunately, we have found this method difficult to implement in
practice. The process of deconvolution naturally pushes
information into high-$n$ and $m$ shapelet coefficients, as shown
in Shapelets~I figure 8. Although the ensuing galaxy
reconstructions are still realistic, information about the overall
galaxy morphology distribution is spread thinly over an increased
number of coefficients. This distribution is no longer
sufficiently well sampled by galaxies in the HDFs for the
smoothing-and-resampling method presented in \S\ref{method} to be
effective.

An alternative solution exists to simulate images with a PSF of
the same size or larger than that of HST. The WFPC2 PSF can be
conveniently maintained throughout the simulations, and the images
convolved again at the end with a second, `difference' kernel.
This kernel is intended to make up the difference between the
original PSF of WFPC2 and that of the new instrument. It can be
obtained by deconvolving the WFPC2 PSF from the new PSF, an
operation performed easily in shapelet space (see Shapelets~II
\S3). An example of this method can be seen in Massey \etal2003.

\section{Advantages and disadvantages of using shapelets} \label{adddisadd}

\subsection{Advantages of shapelets}

Figure~\ref{fig:gals} demonstrates the superb quality of
shapelet-based image reconstruction possible for all galaxy
morphologies. Particularly for spiral or irregular galaxies, we
find the shapelet models superior to those using traditional
radial profiles alone {\it e.g.}~{\tt GALFIT} (Peng \etal2002).
That paper contains plots similar to figure~\ref{fig:gals}; but
with much worse residuals.

There are also many more advantages to using the shapelet parametrization for
image simulations. For example, the truncation in $n_{\rm max}$ produces data
compression by setting a minimum and maximum physical scale of interest (see
discussion in Shapelets~I). The discarded high-$n$ order coefficients contain a
small amount of high spatial frequency information. But because we have ensured
that the reconstruction has been pursued up to an order $n_{\rm max}$ such that
$\chi^2_r\le 1$, we know that the high-frequency remainder is consistent with
noise. Usefully for astronomy, the resolution of a shapelet model is also
greatest near its centre. The compression factor for typical galaxy
morphologies can be as high as $50$ (Shapelets~I). Furthermore, this
compression is achieved through a parameter-independent truncation of a series.
With its complete basis set, shapelets avoid the requirement in {\tt GALFIT} or
{\tt GIM2D} (Simard 1998) to specify in advance the number and type of profiles
for each model. A Karhunen-Lo\'{e}ve decomposition would also require models to
be specified in advance for both the image and the noise. Furthermore, the
orthonormality of the shapelet basis set guarantees a unique and linear
one-to-one mapping from the image plane to the coefficients. This advantage,
and many of shapelets' convenient mathematical properties are lost to methods
using an over-complete basis set such as {\tt Pixon} (Pi\~{n}a \& Puetter
1993).

It is mainly for these convenient mathematical properties that we
choose to model galaxies using shapelets. For example, an object's
orientation is controlled to first order by the phase of the
$a_{22}$ coefficient (corresponding to the position angle of the
object's ellipticity); and its chirality (handedness) by the
relative phase of the $a_{42}$ coefficient. The first can easily
be factored out of the parametrization, so that the ellipticity of
all objects becomes aligned to the horizontal axis. The image can
then be flipped, if necessary, so that the sign of the $a_{42}$
phase is positive and the outer isophotes of all objects twist in
the same anti-clockwise sense. Correlations between remaining
shapelet coefficients $a_{nm}$ are of course maintained in order
to preserve the morphology of the galaxy. Any two well-sampled
objects which are identical apart from their orientation will then
decompose into identical shapelet coefficients. This greatly
increases the sampling density of the galaxy morphology
distribution. Simulated galaxies will later be randomly rerotated
and flipped as they are created.

Shapelets are also designed to be convenient for many aspects of image
manipulation and post-processing. Since the shapelet basis functions
$\chi_{nm}$ are the eigenfunctions of the 2D Quantum Harmonic Oscillator, they
are invariant under Fourier transform up to a phase factor. This renders
convolutions ({\it e.g.}~with a PSF) easy and quick to perform. It also
suggests a well-developed mathematical notation from quantum mechanics.
Convolutions become a bra-ket matrix multiplication (see Shapelets~II).
Translations and rotations, useful for simulating dithered images, are
described to first order by a few applications of $\hat{a}$ and $\hat{a}^\dag$
ladder operators. So too are the distorting shears produced by both optical
aberrations within a telescope and weak gravitational lensing within galaxy
clusters (see Shapelets~I).

\subsection{Disadvantages of shapelets}

There are two main criticisms often levelled at shapelets. The first is that a
Gaussian-Laguerre expansion may not easily capture the extended wings of many
galaxies. After truncation in $n_{\rm max}$, the shapelet basis set is left
incomplete and not ideally matched to typical exponential or de Vaucouleurs
profiles. A demonstration that our algorithm does select sufficiently high
$n_{\rm max}$ is the remarkable match in the concentration index between
shapelet models and real galaxies shown in \S\ref{tests}. In fact, the ability
of a shapelet decomposition to recognise correlations between adjacent pixels
may even enable it to extend further than {\tt SExtractor} into the wings of a
faint object at the threshold of detection, where flux in individual pixels is
lost beneath the noise.

A potentially greater problem for our simulations is the second criticism that
a shapelet decomposition can produce artefacts when it is truncated. Indeed,
any truncated basis set that is complete rather than over-complete will be
subject to spurious residuals that resemble one basis state, due to the
near-cancelling of large positive and negative coefficients in others. For
shapelets, this emerges as ringing, and is particularly apparent after PSF
deconvolution or around long and thin galaxies, which are less well-matched to
the circular basis functions. Furthermore, a desire to keep the shapelet
decomposition method linear prevents the imposition of a positive-definite
constraint. The spurious residuals can therefore appear as either extra
positive flux or negative holes. However, we note that this occurs widely in
other methods, including wavelets, where it is only removed by a (non-linear)
projection in wavelet space onto the sub-space of positive solutions. While
most low-level residuals will be lost in the final simulated images beneath
even modest background noise, we turn around this disadvantage in \S\ref{lam}.
There we use the absence of any negative holes in a noise-free image as a
first-order diagnostic that the morphed shapes of simulated galaxies are
realistic.

\section{Shapelet parameter space} \label{method}

\subsection{The multi-dimensional Hubble Tuning Fork} \label{hsv}

A sample of galaxy morphologies can be thought of as a
distribution of points in a multi-dimensional shape parameter
space. The axes in this space might represent size, magnitude,
position angle (P.A.) and so on. Each point corresponds to a
particular galaxy with a specific morphology, and various
correlations may emerge between variables. For example, the
classic Hubble tuning-fork diagram (Hubble 1926, Sandage 1961, de
Vaucouleurs 1959) relates the object ellipticity, the bulge/disc
ratio, and the extent to which the spiral arms are unwound. {\tt
GIM2D} (Simard 1998) and {\tt GALFIT} (Peng 2002) software use
axes parametrized by the relative amounts of exponential or de
Vaucouleurs/S\'{e}rsic functions (de Vaucouleurs 1959, S\'{e}rsic
1968) required to fit the galaxy's radial profile.

\subsection{The multi-dimensional Shapelet Tuning Fork} \label{polars}

In this work, we instead choose the axes of our galaxy morphology distribution
to be the magnitudes and complex phases of the polar shapelet coefficients.
First, we describe the properties of this `shapelet parameter space'. In the
following section, we will then argue that the underlying Probability Density
Function (PDF) of galaxy morphology is relatively simple in this parameter
space and may be recovered from a finite sample like the HDFs.

\begin{figure}
 \epsfig{figure=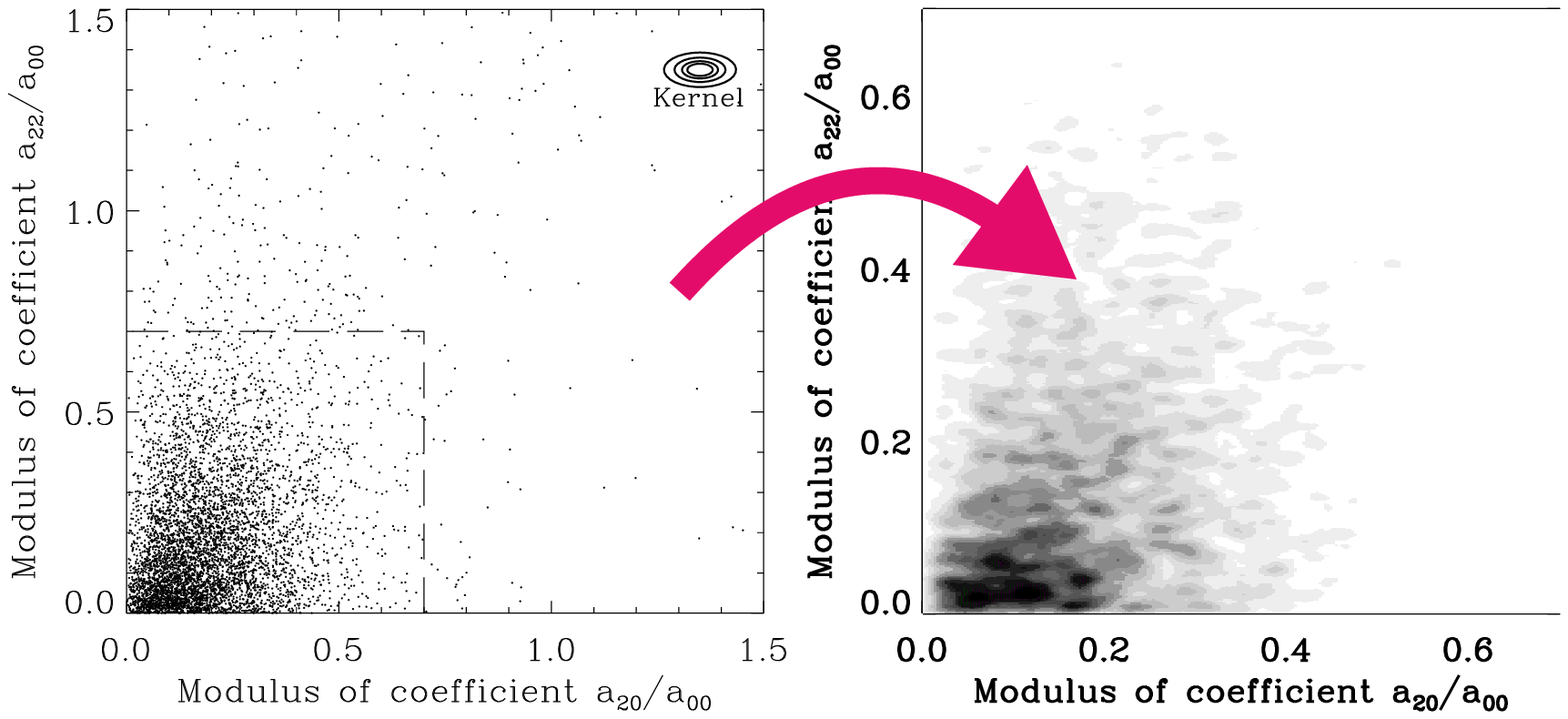,width=84mm}
 \epsfig{figure=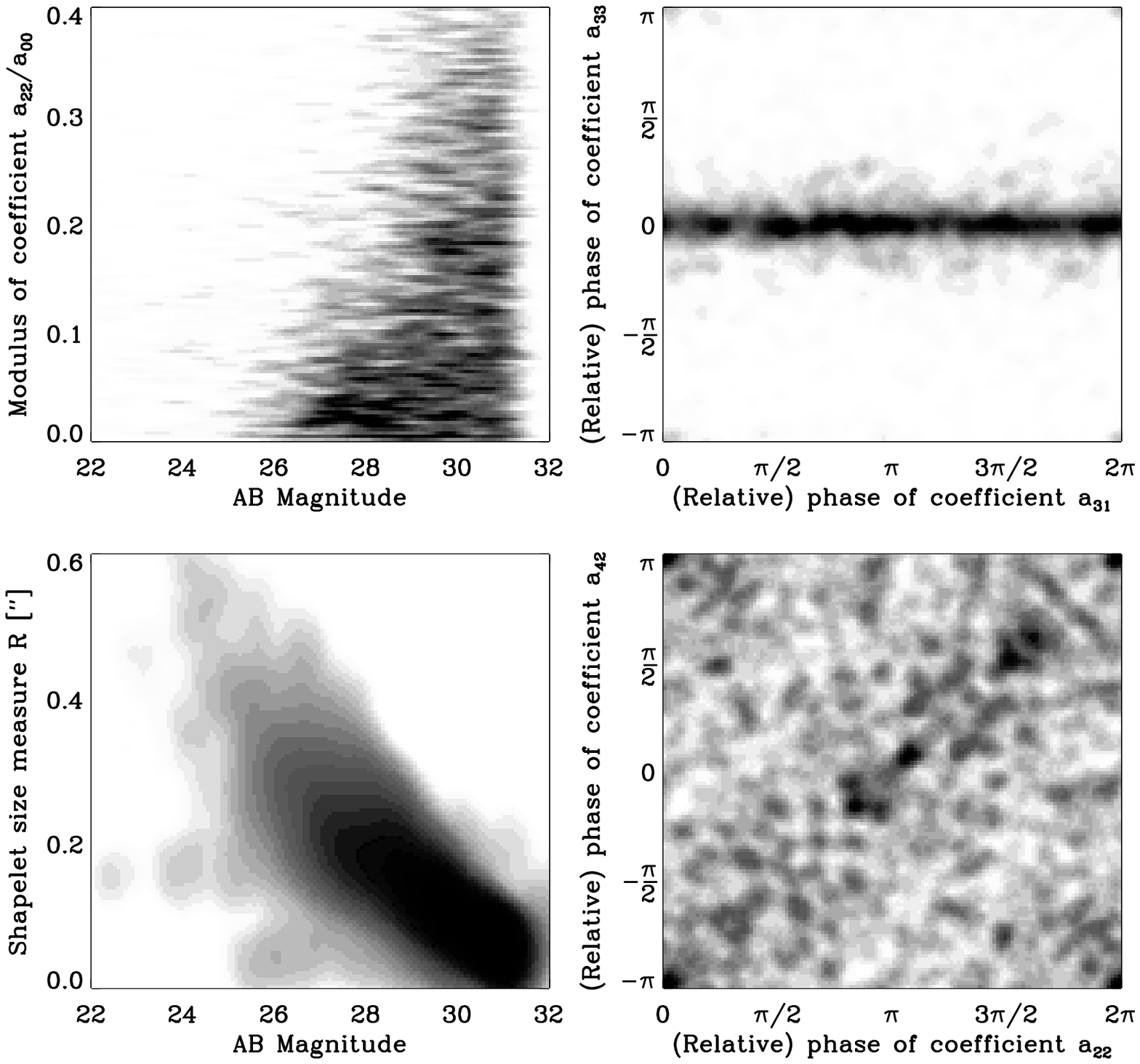,width=84mm}
 \caption{Phase space correlations and smoothing in the shapelet parameter
 space. The top left panel displays the position of measured HDF galaxies along
 two axes of shapelet space; the top right panel shows the probability
 distribution produced by smoothing this distribution. The other left panels
 display further projections of the PDF onto shapelet coefficient, size and
 magnitude axes, while the remaining right panels display phase correlations
 between shapelet coefficients. The colour scale is logarithmic in the bottom
 left panel. \label{fig:slice}}
\end{figure}

Projections of shapelet parameter space are shown in figure~\ref{fig:slice}.
Each point in the top-left panel represents a data vector encoding the shape
information about one HDF galaxy. Collectively, they describe the overall
morphology distribution of distant galaxies. The rotations and reflections used
to pre-align galaxies have considerably compressed this space (without loss of
information) and allowed it to be more densely sampled by only a finite number
of galaxies. Notice that there are correlations evident in the parameters,
which correspond to the construction of the familiar shapes of galaxies. In the
middle-left plot, for example, the scatter of ellipticity values widens for
faint galaxies which are known to be more irregular. In the bottom-right plot,
deviations from the diagonal show twisting isophotes that can grow, with higher
order basis functions, into spiral arms. It is also important to notice that
some regions of parameter space are empty. A random set of shapelet
coefficients will not produce a realistic galaxy shape: there is not even a
positive definite constraint imposed upon an image in the shapelet formalism.

Two other axes are required for our parameter space, since real
galaxy morphologies clearly vary as a function of size and
magnitude ({\it e.g.}~figure~\ref{fig:realimage}). Storing the
shapelet scale factor $\beta$ (see \S\ref{decomp}) allows large
HDF galaxies to occupy different regions of parameter space to
small ones. Similarly, using magnitude as a parameter allows
galaxies of different luminosities to have different shapes. Since
shapelet coefficients (including $a_{00}$) scale as the flux, once
we include magnitude as an independent parameter, we can divide
all $a_{nm}$ by $a_{00}$. This removes explicit magnitude
dependence from these quantities and coincidentally ensures a
convenient version of adaptive smoothing at a later stage (see
\S\ref{pdf_kernel}). The degenerate parameter $a_{00}=1$ is now
removed, and size and magnitude are treated in the same way as any
other axis of the parameter space from now on.

Note that any  orthogonal transformation of the shapelet basis functions would
maintain their useful properties of completeness, orthogonality and Fourier
transform invariance. For instance, the Cartesian version of shapelets can be
used instead (see Shapelets~I), but without the convenient factoring out of the
object's orientation and handedness. Using principal components analysis (PCA;
{\it e.g.} Francis \& Wills 1999), it is possible to calculate the optimal
linear combination of shapelet coefficients to quantitatively describe galaxy
morphology with the fewest numbers. However, both elliptical and spiral galaxy
shapes are already quite simple to manufacture by specifying only a few polar
shapelet coefficients; we therefore avoid the extra complication of PCA in this
paper. Of course, the principle components of galaxy morphology are interesting
in their own right. These are being studied elsewhere.

\subsection{Recovery of a smooth underlying PDF} \label{justify}

The top-left panel of figure~\ref{fig:slice} shows a slice
through the parameter space of galaxy morphologies, populated by
$\delta$-functions representing real, observed shapes. Unlike a
distribution parametrized simply by bulge/disc ratios and disc
inclination angles, it is not obvious {\it a priori} that an
underlying, smooth PDF should exist for galaxy morphologies in
shapelet space. However, the compact shapelet representation of
astronomical objects suggests that this ought to be the case, and
we will attempt to recover it by smoothing this parameter space.

Once the validity of the smoothed PDF has been established, it
will be a simple matter to resample it and thus to
synthesise a population of galaxies. Monte Carlo techniques can be
used to generate unlimited numbers of realistic galaxies in this
fashion, to fill any amount of sky area in a simulated imaging
survey.

The remaining panels of figure~\ref{fig:slice} demonstrate that
the parameter space is indeed smooth in those places where it is
well sampled. We assume that some other regions are equally
smooth, but poorly sampled because of the finite number of
galaxies in the HDF. We note that voids are also expected in the
parameter space, where the shapelet expansions do not correspond
to realistic galaxy shapes. We will therefore be careful not to
smooth the PDF with large smoothing lengths which would
significantly encroach upon these voids. However, limited
perturbations around HDF galaxies may indeed recover realistic
morphologies.

Without an explicitly physical model of galaxy morphology and
evolution built in to shapelets, it is the final results that must
provide the ultimate verification of our statistical method. In
$\S$\ref{results}, we show that it is indeed possible to find a
smoothing length for the PDF that recovers objects which appear to
represent realistic shapes. In $\S$\ref{tests} we demonstrate
quantitatively that their global properties are realistic, by
comparing real and simulated populations of galaxies via
morphology diagnostics commonly used on deep images.

\subsection{Multivariate kernel smoothing method} \label{pdf_kernel}

\begin{figure*}
 \begin{center}
  \begin{tabular}{lr}
   \epsfig{figure=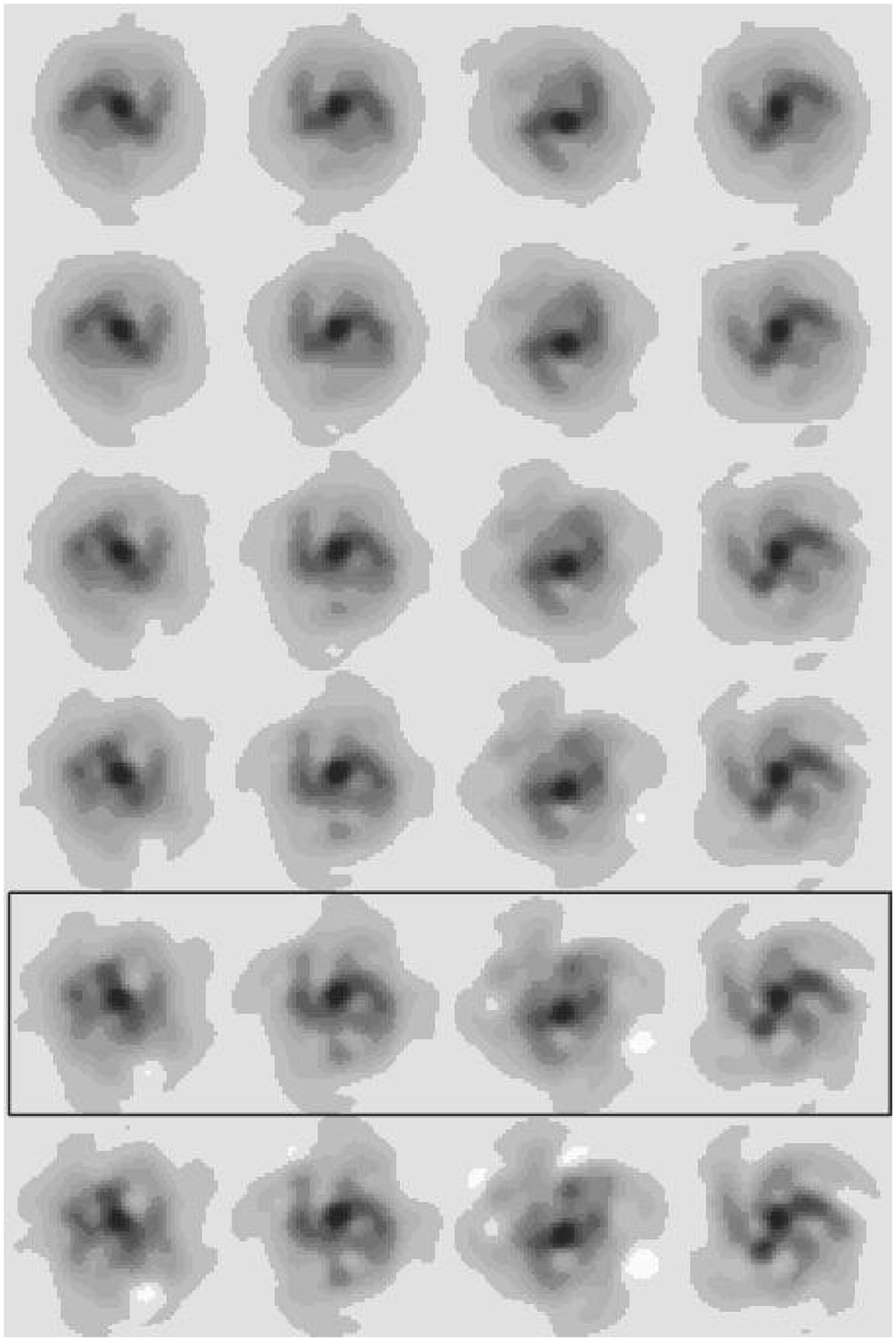,width=84mm} &
   \epsfig{figure=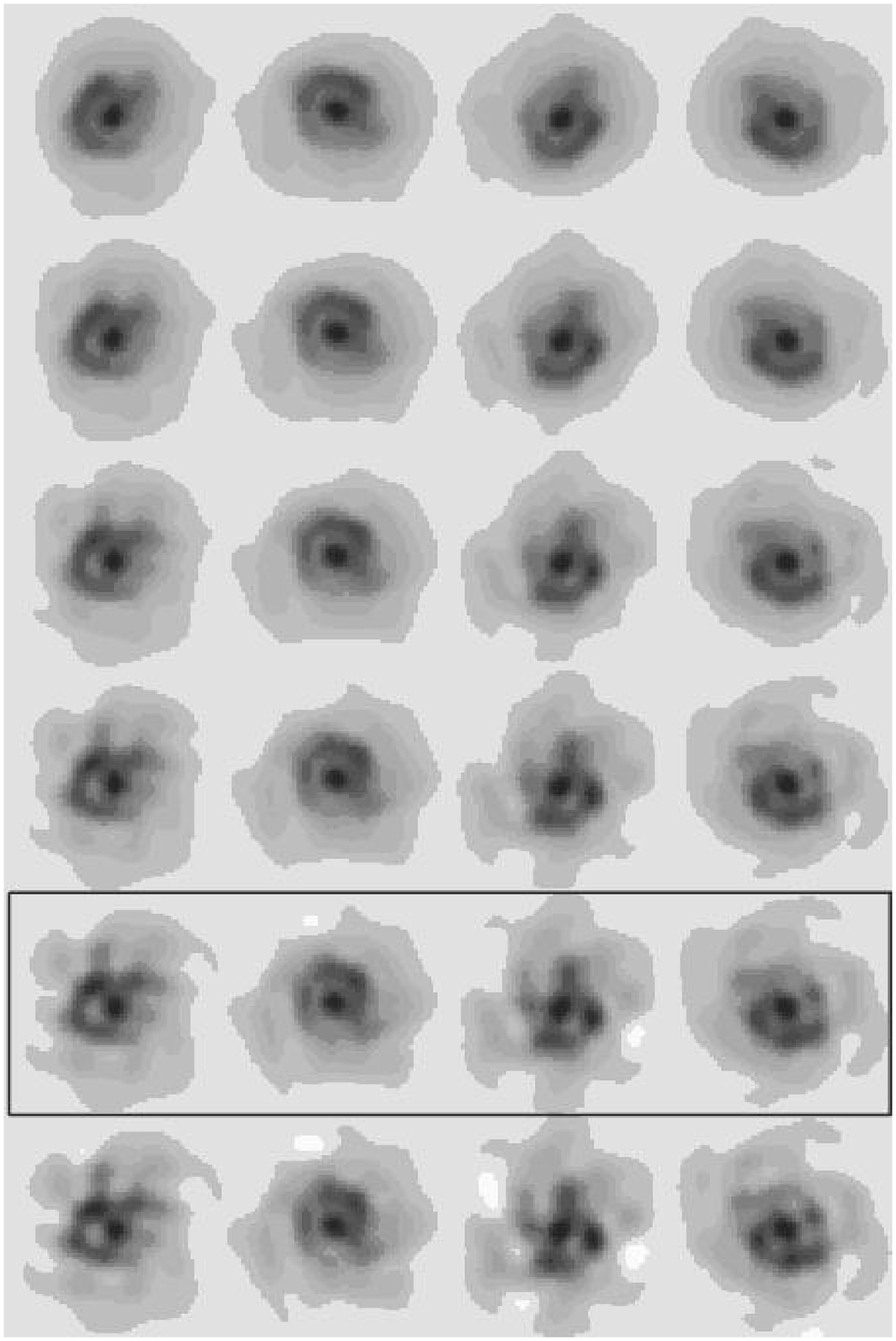,width=84mm} \\
  \end{tabular}
  \caption{The effect of perturbing galaxy morphologies in shapelet
  space. Each image in the top row shows a real HDF galaxy, rotated by
  random angles. Its shapelet coefficients are incrementally perturbed
  in successive rows, although its overall flux is kept constant for the
  purpose of this plot. A degree of perturbation corresponding to our
  choice of the smoothing length $\lambda_i$ is shown inside the box:
  these images represent typical simulated galaxies. Perturbations
  larger than $\lambda_i$ produce objects which contain significantly
  negative pixel values. The left panel depicts a spiral galaxy; the
  right panel a more typical irregular form. The colour scale is
  logarithmic.}
  \label{fig:lam}
 \end{center}
\end{figure*}

\noindent Many practical approaches have been devised to smooth
discrete samplings of a multivariate PDF. Our main constraint in
selecting one of these methods is the very high
dimensionality of our data set. The median $n_{\rm max}$ required
for objects in the HDF is 4. However, even with the efficient data
compression that shapelets can afford, models of the highest S/N
galaxies use values for $n_{\rm max}$ as high as 15. Adding object
size and magnitude, this corresponds to 137 total coefficients,
and this is therefore the maximum number of dimensions required.

To smooth and resample this dataset, we have chosen the Kernel
smoothing method which is eloquently reviewed by Silverman (1986).
Kernel smoothing can be considered as an alternative to using
histograms. It avoids the ambiguity of binning and instead
yields a smooth analytic curve. For 1-dimensional data, each
sample data point is replaced by a smooth Gaussian kernel. To
create a PDF, all the kernels can be summed and then normalised to
integrate to unity. The width of these smoothing Gaussians still
remains to be decided, but methods exist for optimising this
factor. Each kernel can even be given a different width,
calculated as a function of a quick local density estimate, in
order to produce adaptive smoothing.

In data with more than one dimension, each sample point is
replaced by a multivariate kernel. To help overcome the
difficulties associated with the leaking of probability density
into the wings of many-dimensional kernels, we replace the
Gaussian with a more compact {\it Epanechnikov} kernel
(Epanechnikov 1969),
\begin{eqnarray}
 K(\delta x_i)=\left\{
  \begin{array}{cl}
    \frac{3}{4\lambda_i}\left(1-\left(\frac{\delta
    x_i}{\lambda_i}\right)^2\right) & {\mathrm for}~-\lambda_i<\delta
    x_i<\lambda_i \\
    0 & {\mathrm elsewhere,}
  \end{array}\right.
\end{eqnarray}
\noindent where we have reformatted the shapelet coefficients of
each HDF galaxy into a data vector $x_i$, and $\delta x_i$ are
deviations in shapelet space from these real data points. In each
case, $i$ is a coefficient index running from 1 to 137.
$\lambda_i$ are smoothing widths which will be determined for each
direction of our PDF space in \S\ref{lam}. Isodensity contours of
this kernel are multivariate ellipses whose axes are aligned with
those of the co-ordinate axes (see figure \ref{fig:slice}). In
general, they could be allowed to point in any direction (Sain
1999), but we do not find this to be necessary.

We implement an adaptive smoothing of our PDF by reparametrizing
$a_{nm}$ as $\frac{a_{nm}}{a_{00}}$. Given a constant $\lambda_i$,
this creates an effective smoothing kernel for each object of
widths $\lambda_i' = a_{00} \lambda_i$. This functional form is
useful because the brighter HDF objects are less frequent, and are
therefore more isolated in probability space. Since $a_{00}$
roughly correlates to total flux, we obtain a larger smoothing
radius for brighter objects and better recover the underlying
probability distribution. We will prove that this recipe does
produce realistic morphologies in \S\ref{tests}.

\section{Image generation} \label{results}

\subsection{Resampling the galaxy morphology PDF} \label{lam}

Having recovered a realistic and analytic PDF of galaxy morphologies, we now
wish to resample this distribution to generate brand new galaxy populations.
The main advantage of the kernel smoothing approach now becomes apparent.
Without resorting to costly numerical integration, Silverman (1986) $\S$6.4.1
presents a quick bootstrap method to generate a Monte-Carlo sample from a PDF
constructed with $\delta$-functions smoothed by kernels $K$($\delta x$). We
take the following steps to simultaneously smooth and resample the parameter
space of HDF galaxies:
\begin{eqnarray}\label{eqn:sampling}
\left.
\begin{array}{ll}
Step~1: & {\mathrm Randomly~select~one~of~the~original~HDF} \\
        & {\mathrm galaxies,~uniformly~and~with~replacement.} \\
Step~2: & {\mathrm Generate~a~small~perturbation~}\delta x_i~ {\mathrm from~} \\
        & {\mathrm the~probability~density~function~}K(\delta x_i). \\
Step~3: & {\mathrm Add~}\delta x_i~{\mathrm to~the~shapelet~coefficients~}x_i~{\mathrm of}\\
        & {\mathrm the~HDF~galaxy.~This~simulates~a~new} \\
        & {\mathrm galaxy,~sampled~from~the~overall~PDF.}
\end{array}
\right\}
\end{eqnarray}
\noindent This approach is arrived at by simply regarding the PDF
as a sum of small kernels rather than one overall function.
Individually, these kernels are quick to compute; and the
dimensionality of the PDF can even be lowered for faint objects
that require fewer coefficients. The perturbations can be quickly
sampled from an Epanechnikov kernel $K(\delta x)$ by generating
three random numbers from a uniform probability distribution
between $-\lambda_i$ and $\lambda_i$. If the first does not have
the highest absolute value, take it and discard the rest;
otherwise take the second. Iterating this procedure to generate
sufficient objects for a simulated Hubble Deep Field requires only
a few minutes on a 1GHz PC.

We must now decide how to choose the overall smoothing length $\lambda_i$. If
$\lambda_i\equiv 0$, the kernel is a $\delta$-function and the original HDF
objects are recovered exactly. This arrangement will create simulations of
limited practical use, but in \S\ref{tests} they act as an intermediate test of
the shapelet decomposition. As $\lambda_i\rightarrow\infty$, the coefficients
for simulated galaxies become completely random and the objects become
unrealistic. In this limit, since no positive-definite constraint is ever
imposed in the shapelets formalism, we find that simulated objects exhibit
undesirable holes of negative flux. Figure~\ref{fig:lam} shows realisations of
how a typical galaxy from the HDF is altered by increasingly large
perturbations to its shapelet coefficients, showing negative flux for large
$\lambda_i$ perturbations.

\begin{figure}
 \psfig{figure=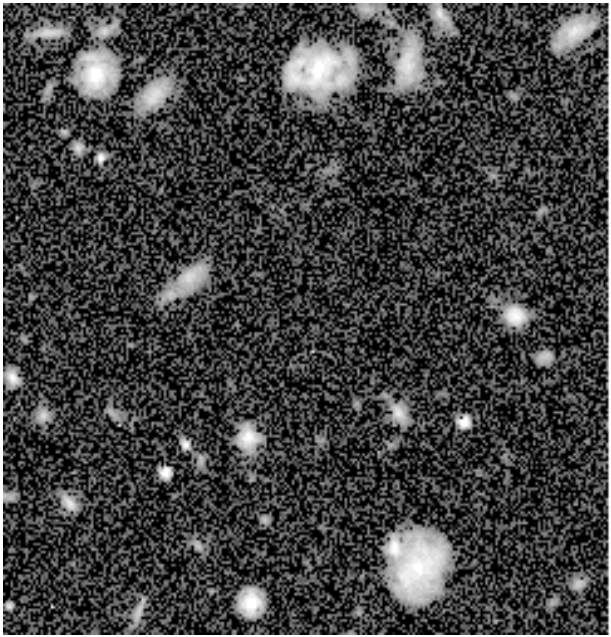,width=84mm,height=84mm}
 \caption{Sample HDF-depth simulated $I$-band image, 30\arcsec$\times$30\arcsec.
 \label{fig:image}}
\end{figure}

We therefore require a choice of $\lambda_i$ which is sufficiently large to
produce new galaxies, yet sufficiently small to maintain realistic
morphological properties. By measuring the minimum pixel values of many
different galaxy realisations, we find suitable results if $\lambda_{phase}\la
15^\circ$ and $\lambda_{moduli}\la 4\times$[mean separation between nearest
neighbours in that dimension]; beyond these values, negative holes rapidly
appear. For the purposes of this paper, we therefore fix $\lambda_i$ to these
limiting values. This still represents relatively weak smoothing, but the
variety and realism of generated morphologies is pleasantly surprising: polar
shapelets are indeed sufficiently close to the Principal Components of galaxy
morphology that small perturbations in shapelet space correspond to reasonable
and realistic changes within galaxy types. A quantitative demonstration of
these remarkable results is presented in \S\ref{tests}.

\subsection{Scattering galaxies on the sky} \label{galdensity}

A Monte Carlo population of genuinely new yet realistic objects has been
extracted from the PDF of galaxy morphology. These galaxies are now allocated
random orientations and locations on the sky, at a density of $700$ per
arcmin$^2$. This constant has been calibrated to recover the same total number
counts, after the addition of noise, as are measured in the HDFs (see
\S\ref{sizemag}). No attempt is made here to correctly model the 2-point
correlation function of galaxy positions, or to include galaxy mergers beyond
those sufficiently advanced to appear as one object in the input {\tt
SExtractor} catalogue.

The correct slope in the size and magnitude distributions are automatically
ensured over a wide range of validity, since size and magnitude are intrinsic
variables of the PDF (see the bottom-left panel of figure~\ref{fig:slice}).
However, it is important to consider the question of completeness in our
simulations for very faint galaxies. A discrepancy could arise through either
non-detections of faint HDF galaxies by {\tt SExtractor} or non-convergence of
their shapelet decompositions. The first effect is minimised by our choice of
{\tt SExtractor} parameters (see \S\ref{catgen}) and the second is shown in
\S\ref{decomp} to be under control. However, the number counts of galaxies at
the very faint end ($I\simgt29$) are also highly sensitive to the the precise
background noise properties (see \S\ref{imgen}). For this reason, we choose not
to consider galaxies fainter than $I=29$.

\begin{figure}
 \psfig{figure=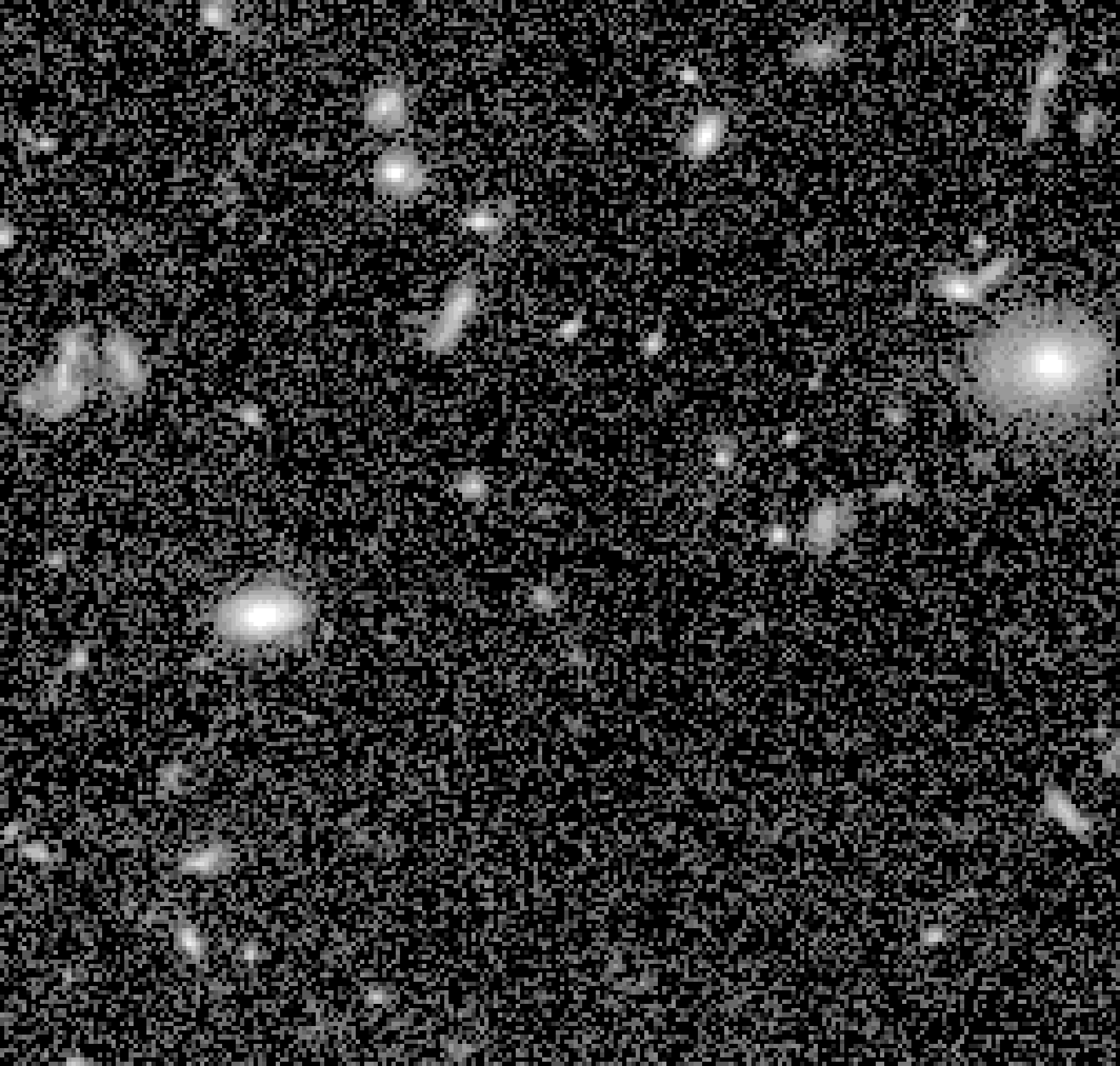,width=84mm,height=84mm}
 \caption{ Section of the real HDF, with the same size and scale as figure
 \ref{fig:image}. \label{fig:realimage}}
\end{figure}

At the bright end, we also expect the simulations to be incomplete, since the
HDFs were intentionally chosen by STScI as areas containing few large, bright
galaxies. In the future, we will extend our simulations in this respect by
incorporating `Groth survey strip' (Groth \etal1994) and ACS galaxies into the
object source catalogue. One could also compensate for any known incompleteness
by preferentially selecting for under-represented galaxy types in step 1 of
procedure (\ref{eqn:sampling}).

\subsection{Modelling telescope and observational effects} \label{imgen}

The shapelet models of galaxy images are actually analytic functions. These can
quickly be convolved with a PSF that has also been decomposed into shapelets,
using the matrix operation in Shapelets~II $\S$3.1. Stars can also be included
in an image, given a magnitude distribution, by repeatedly placing the shapelet
model of the PSF in an image at the appropriate flux amplitude. All of these
analytic objects are then integrated within square pixels of the same
0.0398\arcsec resolution as the {\tt DRIZZLE}d Hubble Deep Field. Our images
have a somewhat larger solid angle than the HDFs because the missing quarter
from the WFPC `L' is restored.

Observational noise can now be added, at a level appropriate to
the desired exposure time. We have simply added photon counting
noise (proportional to the square root of the raw pixel values),
and Gaussian background noise (with an amplitude determined from
the HDF itself). However, it would be easy to add a background
level, cosmic rays and even instrumental distortions: the shearing
for which could be performed conveniently in shapelet space before
pixellisation. A further effect, not included in our simple model,
is noise that is correlated between adjacent pixels. Aliasing occurs
as a side-effect of the {\tt DRIZZLE} algorithm, which recovers
image resolution by stacking several dithered exposures. This
aliasing can make it possible to detect slightly fainter objects
and also introduces some spurious objects at very low S/N. The
steep slope of the real number counts beyond $I=29$ exacerbates this
problem, and we would not yet trust the noise model on our
simulations for galaxies any fainter than this.

Final output is as a FITS image, a sample of which is displayed in
figure~\ref{fig:image}. Larger images may be downloaded via anonymous {\tt ftp}
from the shapelets web page at {\tt
http://www.ast.cam.ac.uk/$\sim$rjm/shapelets}. Notice the wide range of galaxy
morphologies and behaviours present in figure~\ref{fig:image}. In particular,
features resembling spiral arms, dust lanes and resolved knots of star
formation are present, together with various radial profile shapes. By eye, the
simulated galaxies look very similar to those in a similarly-scaled section of
the HDF itself, reproduced in figure~\ref{fig:realimage}. We will
quantitatively examine whether our simulation effectively mimics the morphology
distribution of HDF galaxies in the following section.

\section{Statistical tests and results} \label{tests}

We now demonstrate quantitatively that our simulated images are realistic, in
the sense that commonly used morphology measures for our galaxies match the
distributions of those for galaxies in the HDFs. First, we consider the number
counts and size distributions, using photometry and size measures from {\tt
SExtractor} (Bertin \& Arnouts 1996).These ought to be roughly consistent by
construction, because they are closely related to two of the axes in our
parameter space. Then we compare more detailed morphology measures, such as
concentration (Bershady \etal2000), asymmetry (Conselice \etal2000a),
clumpiness (Conselice \etal in preparation) and ellipticity. These are not
automatically expected to match, because our shapelet-based PDF does not
directly represent these quantities. Thus, a comparison between these
properties for simulated and real data provides a rigorous and fair test of how
realistic our simulations are.

\subsection{Size and magnitude} \label{sizemag}

In order to carry out these tests, we first apply the {\tt
SExtractor} object-finding and shape measurement package on the
version 2 reductions of the HDF-N and HDF-S (Williams \etal 1996,
1998), together with a 6 arcmin$^2$ simulated image of the same
depth. As an intermediate test, we also analyse a simulated image
containing shapelet reconstructions of galaxies drawn from a PDF
left as $\delta$-functions. These should be identical to the
objects in the HDF and act as a test of the shapelets modelling
procedure rather than the perturbations in shapelet space. In all
four cases, approximately 320 galaxies brighter than $I\le29$ were
detected per arcmin$^2$. For the galaxies only, we extracted
observed magnitudes (MAG\_BEST) and sizes (FWHM\_IMAGE).

\begin{figure}
 \psfig{figure=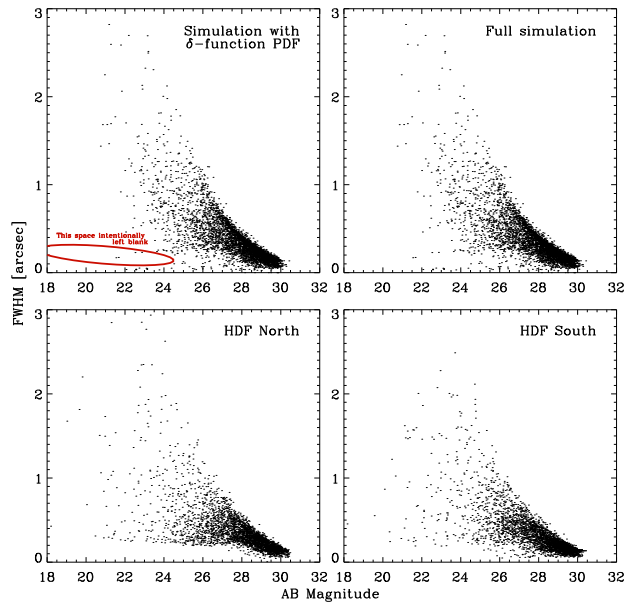,width=82mm,height=82mm}
 \caption{Size {\it vs} magnitude plane for 6 arcmin$^2$ $I$-band
 images to HDF depth, measured with {\tt SExtractor}. 
 Top-left panel: for a simulated image containing shapelet reconstructions
 of HDF galaxies (the PDF kept as $\delta$-functions).
 Top-right panel: for a simulated image with galaxies perturbed in
 shapelet space.
 Bottom panels: for real galaxies in the Hubble Deep Fields North and South,
 calculated using the same {\tt SExtractor} input parameters as reference.
 The stellar locus is omitted from all panels. \label{fig:sizemag}}
\end{figure}

\begin{figure}
 \psfig{figure=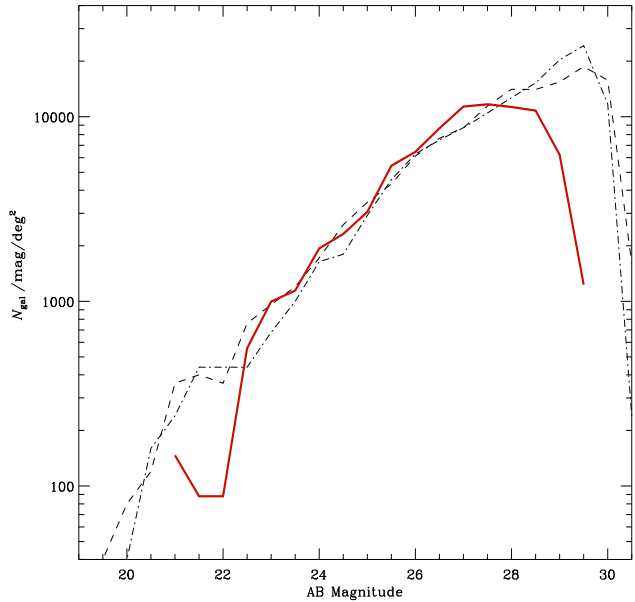,width=84mm,height=84mm}
 \caption{Number counts in simulated $I$-band images (solid red), normalised
 by area on the sky. Also shown are number counts for the Hubble
 Deep Field North (dot-dashed) and South (dashed).}
 \label{fig:numcount}
\end{figure}

Figure~\ref{fig:sizemag} compares the size {\it vs} magnitude distributions of
the simulated images with those of the two HDFs, excluding the stars.
Figure~\ref{fig:numcount} then shows the galaxy number counts for real and
simulated cases in more detail. These match well over six or more orders of
magnitude, whether the simulations used a $\delta$-function PDF or the full
version. Note, however, that the noise in the simulated images is not aliased
in the same way as the {\tt DRIZZLE} algorithm has caused the real data to
become (see \S\ref{imgen}). The number counts beyond $I\sim29$ are highly
sensitive to background noise properties, and are indeed increased in the
simulated image if we artificially smooth the noise. Clearly {\tt DRIZZLE} is
something that needs further attention in a future implementation.

\begin{figure}
 \psfig{figure=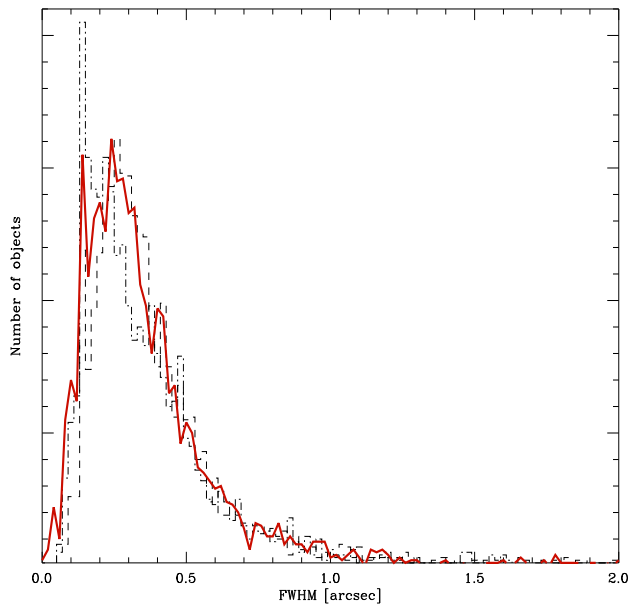,width=84mm,height=84mm} \caption{Size
 distribution of objects in a 6 arcmin$^2$ simulated
 image with limiting magnitude $I=29$ (solid red). Also shown are
 size distributions for the Hubble Deep Field North (dot-dashed)
 and South (dashed).}
 \label{fig:sizedist}
\end{figure}

For the present purposes, we apply magnitude cuts and compare only
the brighter objects, which are unaffected by such minor changes.
These cuts are at levels determined by the stability of an
individual diagnostic to noise. Figure~\ref{fig:sizedist} compares
the size distribution of the simulated objects brighter than
$I=29$ with those of the HDF galaxies, as found by {\tt
SExtractor}. We find that there is excellent agreement in the
shape of this distribution: the median and standard deviation FWHM
for real galaxies in the HDFs are 0.30\arcsec and 0.24\arcsec. For simulated
objects, these figures are 0.31\arcsec and 0.23\arcsec. This agreement comes
about partly (but not entirely) by construction. It was somewhat
expected that our simulated images will closely match real data in
terms of their magnitude and size distributions, but the final
high precision is encouraging.

\subsection{Galaxy morphology diagnostics}\label{morphology}

\begin{figure}
 \psfig{figure=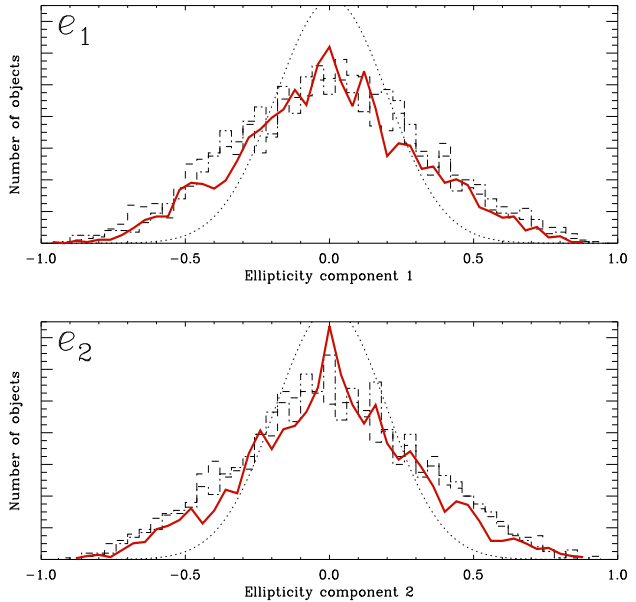,width=84mm,height=84mm}
 \caption{Ellipticity distribution, as defined in equations (\ref{eq:e1}) and
 (\ref{eq:e2}), of objects in 6 arcmin$^2$ simulated image with
 limiting magnitude $I=29$ (solid line). Also shown is the ellipticity
 distribution for the Hubble Deep Fields North (dot-dashed) \& South (dashed),
 and a Gaussian with the same mean and rms (dotted).}
 \label{fig:edist}
\end{figure}

We can more stringently test the reliability of our algorithm to reproduce
properties of real galaxies by measuring morphological parameters which are
entirely independent of shapelets. We apply a series of commonly used
morphology diagnostics to two different realisations of the simulated images. A
first version, containing unaltered shapelet models of HDF galaxies, tests the
shapelet modelling process in isolation. A second simulated image, with
galaxies drawn from the fully smoothed morphology PDF tests the fairness of
these perturbations in shapelet space.

A first basic analysis is to determine the gross shape of galaxies, {\it i.e.}\
their ellipticities. The ellipticity of all the galaxies was obtained from {\tt
SExtractor}. Following a convention in weak lensing literature, we here define
two independent components of ellipticity as

\begin{equation}
  e_1 \equiv
  \frac{{\mathrm A\_IMAGE}^2-{\mathrm B\_IMAGE}^2}
       {{\mathrm A\_IMAGE}^2+{\mathrm B\_IMAGE}^2}
  \cos (2\times{\mathrm THETA\_IMAGE})
  \label{eq:e1}
\end{equation}

\begin{equation}
  e_2 \equiv
  \frac{{\mathrm A\_IMAGE}^2-{\mathrm B\_IMAGE}^2}
       {{\mathrm A\_IMAGE}^2+{\mathrm B\_IMAGE}^2}
  \sin (2\times{\mathrm THETA\_IMAGE})
  \label{eq:e2}
\end{equation}

\noindent where A\_IMAGE and B\_IMAGE are the lengths of the major and minor
axes of the ellipse, and THETA\_IMAGE is the angle between the major axis and
the horizontal (all parameters supplied by {\tt SExtractor}).
Figure~\ref{fig:edist} compares this ellipticity distribution of the real and
fully simulated objects brighter than $I=29$. Again, these are in excellent
agreement: with standard deviations in $e=\sqrt{e_1^2+e_2^2}$ of 0.64 for real
data, 0.62 for simulated data using a $\delta$-function PDF and 0.62 for
simulated data using the full PDF.

\begin{figure}
 \psfig{figure=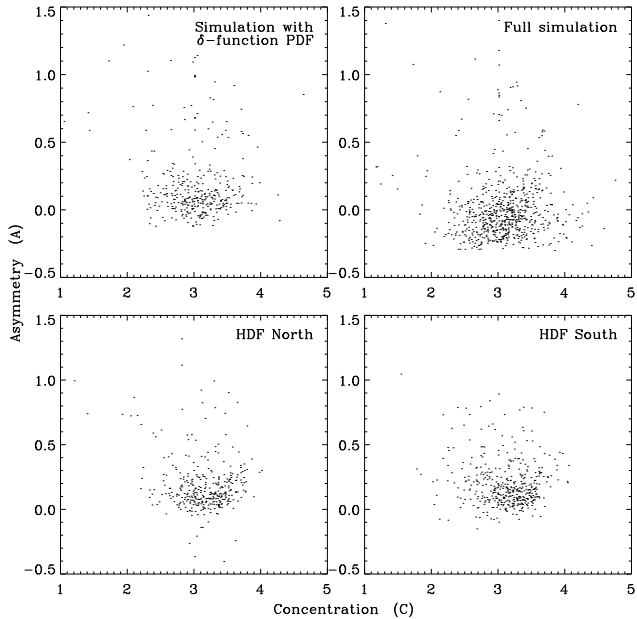,width=84mm}
 \caption{Concentration {\it vs} asymmetry, as defined
 in equations (\ref{eq:c}) and (\ref{eq:a}), for 6 arcmin$^2$ images
 with limiting magnitude $I$ = 26.
 Top-left panel: for a simulated $I$-band image containing shapelet
 reconstructions of HDF galaxies (the PDF kept as $\delta$-functions).
 Top-right panel: for a simulated image with galaxies perturbed in 
 shapelet space.
 Bottom panels: for real galaxies in the Hubble Deep Fields North and 
 South. \label{fig:ca}}
\end{figure}

\begin{figure}
 \psfig{figure=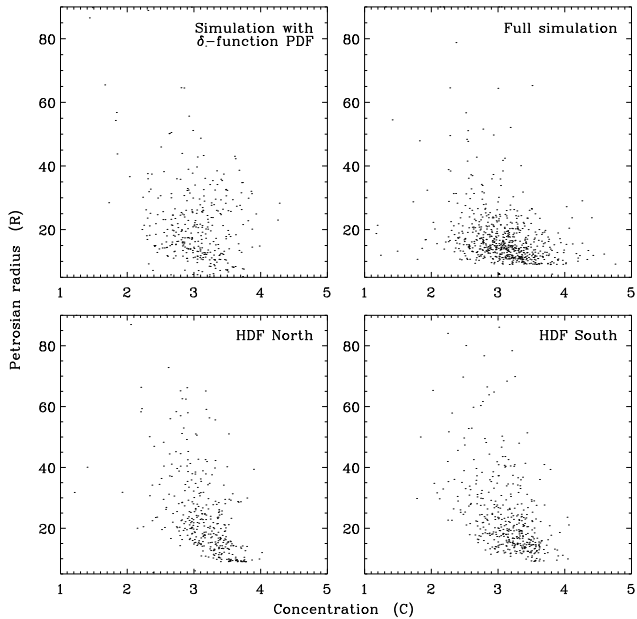,width=84mm}
 \caption{Concentration {\it vs} Petrosian radius, as defined
 in equation (\ref{eq:c}) and the text, for
 6 arcmin$^2$ square images with limiting magnitude $I$ = 26.
 Panels are ordered as in Figure~\ref{fig:ca}. \label{fig:cr}}
\end{figure}

\begin{figure}
 \psfig{figure=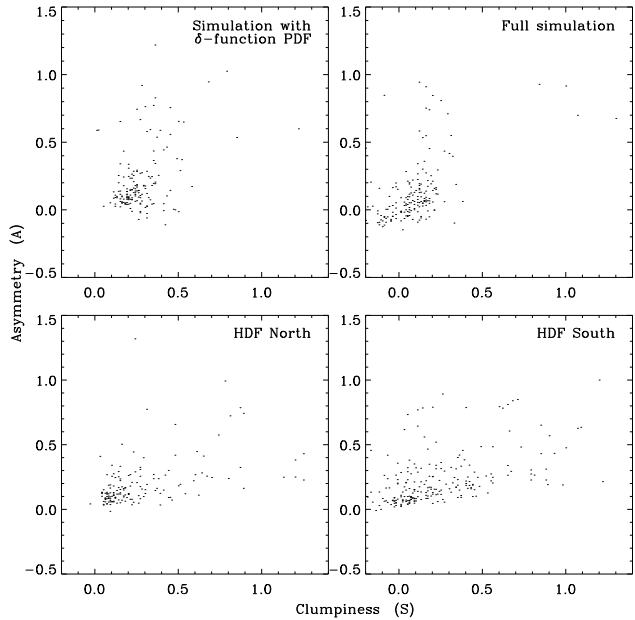,width=84mm}
 \caption{Asymmetry {\it vs} clumpiness, as defined in equations
 (\ref{eq:a}) and (\ref{eq:s}), for 6 arcmin$^2$ square images with
 limiting magnitude $I$ = 25.
 Panels are ordered as in Figure~\ref{fig:ca}. \label{fig:clump}}
\end{figure}

The four images have also been passed through the model-independent morphology
software developed by Conselice \etal(2002a), Bershady \etal(2000) and
Conselice (2003), in order to measure the concentrations ($C$), asymmetries
($A$) and clumpiness ($S$) values of the real and simulated galaxies. We first
describe how these three quantities are calculated, and then compare the
distributions obtained for these measures from real data and simulations. These
`$CAS$' parameters are very informative, as all nearby galaxy types
(ellipticals, spirals, dwarfs, {\it etc.}) fall in distinct regions of $CAS$
space (Conselice 2003). These parameters thus capture most of the variation in
galaxy structures and have frequently been used for quantitative morphology
classification.

The concentration index, $C$, is defined in terms of the ratio of the radii
containing 80\% ($r_{80}$) and 20\% ($r_{20}$) of the object's total flux:

\begin{equation}
  C \equiv 5 \times \log \left( \frac{r_{80}}{r_{20}} \right) ~.
  \label{eq:c}
\end{equation}

\noindent For the total flux, we use the flux within an aperture $1.5$ times
the size of the Petrosian radius at $\eta = 0.2$ (Bershady \etal2000). The
$\eta$ parameter is defined as the ratio of the surface brightness at a radius
divided by the surface brightness integrated within the radius, such that at
the centre of a galaxy, $\eta = 1$ and at the very edge of a galaxy (where its
surface brightness is 0), $\eta = 0$.

Typical values of $C$ for real galaxies range from approximately 2
to 6. Galaxies with $C>4$ are usually ellipticals or spheroidal
systems: a galaxy with an $r^{1/4}$ profile has $C=5.2$. A purely
exponential disc galaxy has $C=2.7$ (Bershady \etal2000). Objects
with lower light concentrations are shown by Graham \etal(2001) to
be systems with low central surface brightnesses and often low
internal velocity dispersions. Low concentration values are also
found for dwarf galaxies ({\it e.g.}~Conselice \etal2002). The
concentration index thus correlates, within some scatter, with the
total mass of a galaxy.

The asymmetry index used in this paper (called $A_{180}$ in Conselice \etal
2000a,b) is calculated by rotating an image by $180\degr$ and subtracting
the it from the original. Then we evaluate

\begin{equation}
  A \equiv
  \min \left[\frac{\sum|I_{x,y}-I_{x,y}^{180}|}{\sum|I_{x,y}|} \right] -
  \min \left[\frac{\sum|B_{x,y}-B_{x,y}^{180}|}{\sum|I_{x,y}|} \right] ,
  \label{eq:a}
\end{equation}

\noindent where $I_{x,y}$ is the galaxy surface brightness in the $(x,y)$ pixel
of the image, $B_{x,y}$ the sky background in the same pixel, and superscripts
denote rotations. Sums are over all pixels within the same $\eta=0.2$ Petrosian
radius from which the total light measurement is made. Minimisation is then
over different choices of the centre of rotation (see Conselice \etal2000a).

The asymmetry index is sensitive to any physical processes in a galaxy that
produce asymmetries in light distributions, such as star-formation, galaxy
interactions/mergers, and projection effects such as dust lanes. There is a
general correlation between the asymmetry value and the ($B-V$) colour
(Conselice \etal2000a). Since most galaxies are not edge-on systems, star
formation and galaxy interactions/mergers are the dominant effects that produce
asymmetries in real galaxies. These two effects can often be distinguished,
however. Systems with asymmetries $A>0.35$ are generally created by
interactions or mergers (Conselice 2003, Conselice \etal2003). However, other
merger events can have more modest asymmetry values. From this and more
detailed studies of the asymmetry index, it has been concluded that $A$ is most
sensitive to bulk structures in galaxies (Conselice 2003).

The clumpiness parameter, $S$, is a measure of the high-spatial frequency
component of galaxies. It is calculated by smoothing a galaxy's image with a
smoothing length $\sigma$, then subtracting this smoothed version
$I_{x,y}^{\sigma}$ from the original image. This leaves a residual map
containing only those features with a high-spatial frequency. Summation is
again performed over pixels within the $\eta=0.2$ Petrosian radius, although
those from the central cusp are ignored. Also including a correction for the
background B$_{\rm x,y}$, the clumpiness is then defined as

\begin{equation}
  S \equiv 10 \times
  \frac{\sum_{xy}|I_{x,y}-I_{x,y}^{\sigma}|
      - \sum_{xy}|B_{x,y}-B_{x,y}^{\sigma}|}
       {\sum_{xy}I_{x,y}} ~.
  \label{eq:s}
\end{equation}

\noindent The clumpiness index is sensitive to the instantaneous rate of star
formation, and correlates very well with H$\alpha$ equivalent widths; it also
correlates to a lesser degree with broad-band colors (Conselice 2003). Other
details of its calculation and properties are discussed in detail in Conselice
(2003).

We also use the Petrosian radius $R$ (Petrosian 1976) to characterise the
galaxies, defined as the position where $\eta=0.2$. The Petrosian radius is
found to be a better index than the {\tt SExtractor} FWHM radius for
determining morphological sizes, as {\tt SExtractor} radii are based on
isophotal thresholds which will represent different physical distances from the
galactic centre depending on the distance to the galaxy. Because $\eta$ is a
ratio of surface brightnesses in a given galaxy, the run of $\eta$ with $r$ in
a galaxy is immune to many such types of systematic effects (Sandage \&
Perlmutter 1990) and Petrosian radii are found to be a stable tool for deriving
morphological parameters independent of distance (Bershady \etal2000).

We are now in a position to compare the measurements for $C$, $A$,
$S$ and $R$ for real and simulated images. Projections from this
morphological parameter space for real and simulated data are
displayed in figures~\ref{fig:ca}--\ref{fig:clump},
and relevant statistics are compiled in table \ref{tab:stats}.

\begin{table}
 \begin{center}
  \begin{tabular}{cllll}
   \hline \hline
               & HDF-N & HDF-S & Simulation with    & Full \\
   ~~~~~~~~~~~ &       &       & $\delta$-function PDF & simulation \\
   \hline
   $\langle C \rangle$ & 3.11 & 3.13 & 3.03 & 3.07 \\
   rms $C$             & 0.39 & 0.40 & 0.44 & 0.42 \\
   $\langle A \rangle$ & 0.18 & 0.17 & 0.19 & 0.07 \\
   rms $A$             & 0.20 & 0.22 & 0.27 & 0.25 \\
   $\langle S \rangle$ & 0.23 & 0.28 & 0.27 & 0.08 \\
   rms $S$             & 0.28 & 0.28 & 0.15 & 0.19 \\
   rms $e$             & 0.64 & 0.64 & 0.62 & 0.62 \\
   \hline \hline
  \end{tabular}
 \end{center}
 \caption{Galaxy morphology statistics. The first two columns show
 results for real objects, taken from the Hubble Deep Fields.
 Compare this with objects in simulations created using a
 $\delta$-function PDF or the full shapelet-morphing procedure.}
 \label{tab:stats}
\end{table}

As can be seen from the scatter in the plots, the agreement between simulations
and real data is rather good: we are very pleased by the encouraging results.
The matching distributions of the concentration parameter puts to rest one
criticism frequently levelled at shapelets (see \S\ref{adddisadd}), that a
truncated Gaussian-Laguerre expansion may not stretch far enough spatially to
capture the extended wings of typical astronomical objects. Clearly our
algorithm sets $n_{\rm max}$ high enough to avoid this problem while still
modelling the HDF galaxies using only a few coefficients.

The final population of simulated galaxies does contain asymmetry values lower
than those in the real data, although the distributions agree within 1$\sigma$.
This slight discrepancy is neither due to deficiencies in the shapelet
modelling procedure, nor to the increased clustering of galaxies at short
separations in real data, because it is absent from the simulation created with
a $\delta$-function PDF. Decreased object asymmetry must therefore be a
by-product of the PDF smoothing. There is no obvious {\it a priori} reason why
this should happen. Even $m$ states are symmetric and odd $m$ states
anti-symmetric, so if the absolute values of all coefficients are randomly
changed by the same amount, the overall symmetry of the object should stay
constant. However, our nearest-neighbour prescription from \S\ref{lam} results
in an average smoothing length across typical even $m$ states, and particularly
the $m=0$ states, of approximately twice that for odd $m$ states. This may
simply be because the first state is even, and the smoothing length tends to
get shorter as $n$ increases. A more sophisticated adaptive smoothing method
might be found to prevent this effect, but we have not pursued that idea here.
We note the asymmetry discrepancy, but note also that it is relatively small.

The behaviour of the clumpiness parameter is also reasonable. Truncation in
shapelet modelling smooths galaxies slightly, and thus removes the tail of
objects with very high $S$. Morphing in shapelet space apparently acts to then
smooth some of the galaxies further. This is peculiar because, if anything, the
galaxies in figure~\ref{fig:lam} appear by eye to become more clumpy as the
smoothing length is increased. Overall, the agreement of the simulated
distributions with real data is remarkably consistent with the field-to-field
variation between the two HDFs. Indeed, clumpiness is a rather unstable
statistic to measure. For example, even the slight rise in mean clumpiness for
the $\delta$-function simulation might be significant: especially since it is
apparent despite the missing tail at high $S$. It is possible that the increase
is caused by residual artefacts in the shapelet models, but more plausibly
because the noise in our simulated images is not correlated between adjacent
pixels. The HDFs themselves have been {\tt DRIZZLE}d in order to achieve their
high resolution, a process which also aliases the image. As a simple
approximation to this effect, we have tried smoothing the noise slightly in our
simulations, by a top hat kernel 3 pixels wide. This process does indeed remove
the slight disparity observed in the simulated clumpiness distribution, but
simultaneously creates many false detections of faint, circular objects from
the noise at the magnitude limit around $I\ge29$.

Therefore we conclude that our shapelet simulations obtain similar morphology
distributions to those found in real data. This is most encouraging as these
were not arranged by construction, and the level of realism seen here is a
strong vindication of the shapelet modelling of galaxies. Perturbing shapelet
parameters to create new galaxies can introduce a few minor deviations, but
these are small compared to natural variation between objects, and are well
understood and quantified. We can therefore use shapelets as a tool for
investigating galaxy morphology and for creating realistic simulated images.

\section{Comparison to other methods} \label{compare}

There have been many packages in the literature which simulate astronomical
observations, including {\tt Skymaker} (see Erben \etal2001) and {\tt artdata}
in {\tt IRAF} (Tody 1993). These typically parametrize galaxy shapes using
simple physical models such as ellipses with de Vaucouleurs or exponential
profiles. The smooth variation allowed for these parameters enables them to
generate an unlimited number of unique simulated galaxies. These methods are
particularly valuable for simulating images from ground-based telescopes.
Unfortunately, deep images from HST contain galaxies with resolved features
more complex than these analytical models, so such simulations are useful in
only a limited regime.

This was realised by Bouwens, Broadhurst \& Silk (1998), who designed
simulations to investigate the evolution of galaxy morphology in the HDF.
Indeed, their work succeeds in ruling out pure luminosity evolution of
galaxies: which precisely demonstrates the need for deep image simulations to
contain more irregular and asymmetric morphologies. Their method repeatedly
places the few brightest HDF galaxies onto a simulated image, and is similar to
that which ours would have been, had we left the PDF as an (unsmoothed) sum of
$\delta$-functions. Some physics can be added to rescale and redshift these few
sources, but it remains a very small population from which to simulate a large
imaging survey, and containing members drawn exclusively from the local
universe. Creating realistic images was not the intention of Bouwens,
Broadhurst \& Silk (1998) and, for our objectives, their method would require
the addition of more physics ({\it e.g.}~galaxy evolution, star formation
histories, redshift distributions, {\it etc.}).

Our technique attempts to capture the best aspects of both methods, by defining
a smooth parameter space that can yield an unlimited number of unique galaxies,
but also contains a rich diversity of their morphologies (potentially any
morphology, in fact, since the set of shapelet basis functions is complete).
Since the parameter space is populated via statistical rather than physical
arguments, it is the many tests to which we have subjected our simulated images
that demonstrate the validity of our method. We find a regime spanning six
orders of magnitude in luminosity where our simulations are valid, and their
statistical properties match those of real data. This ability to produce
simulated images containing galaxies with realistic morphologies is a
significant advance.

\section{Conclusions} \label{conclusions}

We have presented a method for generating simulated deep sky images of an
arbitrarily large survey area, as might be observed with extended observations
with the Hubble Space Telescope. These simulated images are populated with all
morphological types of galaxies, based upon those in the Hubble Deep Fields.

The simulated galaxies are drawn from a multi-variate distribution of realistic
morphologies, described using the shapelet formalism (Refregier 2003). In order
to generate this morphology distribution, we decompose all HDF galaxies into
shapelet components using least-squares fitting. We optimise this decomposition
by finding the scale length $\beta$ and number of modes $n_{\rm max}$ which
produces a best shapelet coefficient fit to each galaxy. The resulting
coefficients of HDF galaxies form a cloud of points in shapelet space; these
points are replaced by smooth kernels in order to recover the underlying
probability distribution of real galaxy morphologies. The smooth distribution
is then resampled, using an unbiased Monte-Carlo technique, to obtain new
galaxies.

We place these simulated galaxies onto HDF-sized images, simultaneously
including effects such as PSF, pixellisation, photon shot noise and Gaussian
background noise. The level of detail in the resulting simulated galaxies
includes features such as realistic radial profiles, spiral arms, dust lanes
and resolved knots of star formation.

We have noted that the global morphological properties of the simulated galaxy
population must match those of real galaxies if our simulations are to be
useful. We have demonstrated that this is the case by comparing various
morphology diagnostics in simulated and real galaxies, including number counts
as a function of magnitude, the size distribution, ellipticity distribution,
and concentration, asymmetry and clumpiness indices. A test involving purely
the shapelet decomposition and reproduction of the HDF galaxies preserves all
of these statistics with high precision, and we conclude that a shapelet
decomposition can successfully capture the morphological properties of all
galaxy types. A few slight discrepancies are introduced to the statistics by
perturbing their shapelet coefficients (or smoothing the morphology
distribution) to manufacture genuinely new galaxy shapes. However, these
differences are small compared to even the natural variations between objects.
Several minor effects have been well quantified by our various tests, and their
causes understood for correction in future implementations.

An important application for our simulated images is presented in Massey
\etal(2003), where they are used to predict the sensitivity to weak
gravitational lensing of the proposed SNAP satellite. However, the simulations
presented here are in no way specific to gravitational lensing, and may be used
for testing image analysis in various branches of astronomy. Further simulated
images and catalogues are available from the authors.

A useful extension to this work will be to include `Groth survey strip' (Groth
\etal1994) galaxies and ACS data when constructing the morphology probability
distribution. This will provide future simulations with a more extensive sample
of large, bright galaxies, improving the fidelity of the simulations in this
region of parameter space. A method is also in development to generate
multi-colour simulated images using several HDF passbands and photometric
redshifts. A by-product of this work is a complete morphological catalogue of
all the HDF galaxies in shapelet space. This catalogue will be used in a future
paper on the automated morphological classification of galaxies at high
redshift.

\section*{Acknowledgements}

The authors thank the Raymond and Beverly Sackler fund for travel support. AR
was supported in Cambridge by a PPARC advanced fellowship. DJB was supported by
a PPARC fellowship. We would also like to thank Tzu-Ching Chang for her help in
developing the shapelets method and code. That code would run much slower
without Sarah Bridle and Phil Marshall's insightful statistical trickery.
Thanks to Jason Rhodes for help implementing the morphology tests. We are also
grateful to Richard Ellis, Josh Frieman, Andy Fruchter, Eric Gawiser, Jean-Paul
Kneib and Jean-Luc Starck for ideas, comments and enthusiasm throughout this
work. An anonymous referee provided several insights and ideas which have
improved this paper.

\bsp

\label{lastpage}

\end{document}